\documentclass[twocolumn,showpacs,preprintnumbers,amsmath,amssymb,prl]{revtex4-1}

\usepackage{graphicx}
\usepackage{dcolumn}
\usepackage{bm}
\usepackage{epsfig}
\newcommand{\be}{\begin{eqnarray}}
\newcommand{\ee}{\end{eqnarray}}

\begin{document}

\title{Quantum criticality of spin-1 bosons  in a 1D harmonic trap}

\author{C. C. N. Kuhn$^{1,2}$,   X. W. Guan$^{2}$, A. Foerster$^{1}$, and M. T. Batchelor$^{2,3}$ }
 \affiliation{$^{1}$Instituto de Fisica da UFRGS, Av. Bento Goncalves 9500, Porto Alegre, RS, Brazil\\
$^{2}$Department of Theoretical Physics, Research School of Physics and Engineering,Australian National University\\
$^{3}$ Mathematical Sciences Institute, Australian National University, Canberra ACT 0200, Australia}


\date{\today}

\begin{abstract}

We investigate universal thermodynamics and quantum criticality of spin-1 bosons with strongly
repulsive density-density and antiferromagnetic spin-exchange interactions in a one-dimensional harmonic
trap. From the equation of state, we find that a partially-polarized core is surrounded by two
wings composed of either spin-singlet pairs or a fully spin-aligned Tonks-Girardeau gas depending on
the polarization. We describe how the scaling behaviour of density profiles can reveal the universal
nature of quantum criticality and map out the quantum phase diagram. We further show that at
quantum criticality the dynamical critical exponent $z = 2$ and correlation length exponent $\nu=1/2$.
This reveals a subtle resemblance to the physics of the spin-1/2 attractive Fermi gas.
\end{abstract}

\pacs{03.75.Ss, 03.75.Hh, 02.30.Ik, 05.30.Fk}

\keywords{}

\maketitle

Alkali bosons with hyperfine spins  in an optical trap provide  exciting  opportunities to simulate a 
variety of macroscopic quantum phenomena.  In the spinor gas,  the spin-dipolar collisions significantly change 
the spin states producing rich Zeeman effects. In particular,  
spinor Bose gases with  density-density interaction  and  antiferromagnetic spin-exchange interaction \cite{Ho,Ohmi} 
exhibit  various phases of strongly correlated quantum liquids and are 
thus particularly valuable to study quantum magnetism and criticality.  
The experimental study of quantum criticality and universal scaling behaviour has 
recently been initiated in low-dimensional cold atomic matter \cite{Exp10,Exp11}. 
These advances build on 
theoretical schemes for mapping out quantum criticality in cold atom systems \cite{Campostrini, Zhou-Ho,Erich}. 
In this framework, exactly solved models of cold atoms, exhibiting 
quantum phase transitions,  provide  a rigorous way to  investigate quantum criticality \cite{Guan}.

\begin{figure}[t]
{{\includegraphics [width=0.99\linewidth]{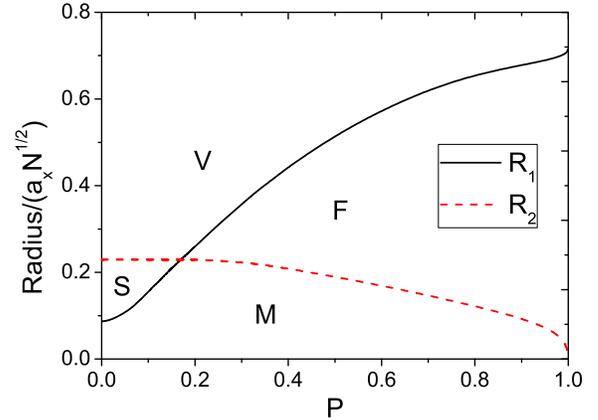}}}
\caption{(Color online) Phase diagram of the spin-1 Bose gas at $T=0$ in terms of  the rescaled axial 
radius $R/a_xN^{1/2}$ versus the polarization $P=N_1/N$ for $N/(a^2_xc^2)=0.05$.  
The lines $R_1$ and $R_2$ denote the phase boundaries  of   vanishing density of singlet-pairs  and  vanishing 
density of  spin-aligned  bosons.  The model exhibits three quantum phases:  spin-singlet paired bosons $S$,  
ferromagnetic spin-aligned bosons $F$, and  a mixed phase of the  pairs 
and unpaired bosons.  $V$ stands for  the vacuum.} 
\label{diagram}
\end{figure}

One-dimensional (1D) spinor Bose gases with short-range delta-function  interaction and antiferromagnetic spin-exchange interaction 
are particularly interesting due to the existence of 
various phases of quantum liquids associated with exact Bethe ansatz solutions \cite{Cao,Shlyapnikov,Lee,Shlyapnikov2}.  
The antiferromagnetic interaction leads to an effective attraction 
in the spin-singlet channel that gives rise to a quasi-condensate of singlet bosonic pairs when  
the external field is less than a lower critical field at zero temperature. In this phase, the 
low energy physics can be characterized  by a spin-charge separation theory of the $U(1)$ 
Tomonaga-Luttinger liquid (TLL)  describing the charge sector  and a $O(3)$ non-linear sigma model 
describing the spin sector \cite{Shlyapnikov}. However,  if the  external field exceeds an upper critical field, 
we have a solely ferromagnetic quasi-condensate of unpaired bosons  with 
spins aligned along the external field.  For an intermediate  magnetic field, the spin-singlet pairs  and 
spin-aligned bosons form a two-component TLL  with a magnetization, see Fig. \ref{diagram}.

Quantum critical phenomena in the spin-1 Bose gas associated with phase transitions at zero temperature
 can be explored by varying the  magnetic field and chemical potential. 
In this paper we use the exact thermodynamic Bethe ansatz (TBA) solution 
for spin-1 bosons  to illustrate the microscopic origin of the quantum criticality of different spin states 
in a 1D harmonic trap. 

{\it The model and equation of state.}- We consider $N$ particles of mass $m$ confined in 1D to a length $L$ with 
$\delta$-interacting type density-density and spin-exchange interactions between two atoms.  The 
Hamiltonian is  \cite{Ho,Cao}
\begin{eqnarray}
\! \! \!  {\cal H}=-\sum^N_{i=1}\frac{\partial^2}{\partial x^2_i}+\sum_{i<j}[c_0+c_2S_i\cdot S_j]\delta(x_i-x_j)+E_z,
\label{Ham}
\end{eqnarray}
where  $S_i$ is a spin-1 operator with $z$-component $(s=1, 0, -1)$, $E_z=-HS^z$ is the Zeeman energy, 
with $H$ the external field and $S^z$ the total spin in the $z$-component. 
The 1D interaction parameters  $c_0=(g_0+2g_2)/3$ and $c_2=(g_2-g_0)/3$ with  
$g_S=-2\hbar^2/ma_S$. 
Here $m$ is the particle mass and $a_S$ represents the 1D $s$-wave scattering length in the total spin $S=0,2$ channels. 
Following  \cite{olshanii,granger} the effective 1D coupling constant $g_S$ can be expressed in terms of the known 
3D scattering lengths.  

Extending Yang's method \cite{Yang},  Cao {\em et al.}  \cite{Cao} solved the model  (\ref{Ham})   for the case $c=c_0=c_2$. 
For convenience, we define the linear density $n=N/L$
and use dimensionless quantities 
$\tilde{\mu}=\mu/\varepsilon_b$, $h=H/\varepsilon_b$, $t=T/\varepsilon_b$
and $\tilde{p}=p/|c|\varepsilon_b$.  Here energy and length are measured in units of 
the binding energy $\varepsilon_b=\hbar^2c^2/16m$ and $c^{-1}$, respectively. We set  $\hbar=2m=1$ in the following.

The model  (\ref{Ham}) has two conserved quantities, 
$S^z$ and the total particle number $N$. 
The number of particles in a particular spin state $(s= 1, 0, -1)$ is no longer conserved.  
For instance the scattering between two particles of spin $s=\pm 1$  can produce two particles of spin $s=0$ and vice versa.
Therefore, the model can have two types of pairing between two spin $\pm 1$ atoms or  between two spin-$0$ atoms, see \cite{Ho,Cao,Lee}.

The ground state properties of the system were studied in \cite{Shlyapnikov,Lee,Shlyapnikov2}. 
Analytic results can be extracted from the TBA equations for strong coupling $\gamma\gg1$ and low temperature $t\ll 1$ (see \cite{kuhn} for details).    
The pressure of the system can be written as $\tilde{p}=\tilde{p}_1+\tilde{p}_2$  in terms of the 
pressure $\tilde{p}_1$  for  unpaired and $\tilde{p}_2$  for  paired  bosons (Boltzmann constant $k_B=1$) where 
\begin{eqnarray}
\tilde{p}_1=-\frac{t^{3/2}}{4\sqrt{2\pi}}f_{\frac{3}{2}}^{1}\left(1+ \frac{t^{3/2}}{32\sqrt{2\pi}}f_{\frac{3}{2}}^{1}-
\frac{124t^{3/2}}{125\sqrt{\pi}}f_{\frac{3}{2}}^{2}\right), \nonumber \\ 
\tilde{p}_2=-\frac{t^{3/2}}{4\sqrt{\pi}}f_{\frac{3}{2}}^{2}\left(1-\frac{124t^{3/2}}{125\sqrt{2\pi}}f_{\frac{3}{2}}^{1}-
\frac{181t^{3/2}}{3456\sqrt{\pi}}f_{\frac{3}{2}}^{2}\right).
\label{pressure}
\end{eqnarray}
Here 
$f^{j}_{n}=Li_n(-e^{A_j/t})$ with $j=1$,$2$ denotes the standard  polylog 
function $Li_n(x)$, with
\begin{eqnarray}
A_1 &=& \tilde{\mu}+h+2\tilde{p}_1-\frac{16\tilde{p}_2}{5}\nonumber \\ 
&+& \frac{t^{5/2}}{16\sqrt{2}}\left( \frac{1}{2\sqrt{\pi}}f_{\frac{5}{2}}^{1}
-\frac{1984}{125\sqrt{2\pi}}f_{\frac{5}{2}}^{2}\right) + f_s, \nonumber \\ 
A_2&=& 1+2\tilde{\mu} -\frac{32\tilde{p}_1}{5}-\frac{\tilde{p}_2}{3} \nonumber \\ 
&-& \frac{t^{5/2}}{16\sqrt{2}}\left( 
\frac{3968}{125\sqrt{\pi}}f_{\frac{5}{2}}^{1} +\frac{181}{108\sqrt{2\pi}}f_{\frac{5}{2}}^{2}\right).
\label{a1a2}
\end{eqnarray}
The terms  $f_s=te^{-h/t}e^{-2\tilde{p}_1/t}I_0(2\tilde{p}_1/t)$ with $I_0(z)=\sum^{\infty}_{k=0}\frac{(z/2)^{2k}}{k!^2}$ are extracted from 
 the spin wave bound states \cite{kuhn}.

 The above result for  the pressure serves as  the equation of state, from 
which thermodynamic  quantities such as the particle density $n$, 
the density of unpaired bosons $n_1$,  the density of spin-singlet pairs $n_2$, the entropy,  and the compressibility of the system can be determined
through the standard  thermodynamic relations. Moreover, the equation of state allows an exact description of TLLs and  
 quantum criticality of the system.

{\it Phase diagram and density profiles.}-
In order to study   quantum criticality in the spin-1 Bose gas in a 1D harmonic trap,  the equation of state (\ref{pressure}) 
can be reformulated within the local 
density approximation (LDA)  by a replacement $\mu(x)=\mu(0)-\frac{1}{2}m\omega_x^2x^2$ 
in which $x$ is the position and $\omega_x$ is the trapping  frequency, see \cite{Yin}.
Using the dimensionless chemical potential  $\tilde{\mu}(\tilde{x})=\tilde{\mu}(0)-8\tilde{x}^2$, where 
$\tilde{x}=x/a_x^2|c|$ is a rescaled distance  and $a_x=\sqrt{\hbar/m\omega_x}$ is the harmonic characteristic radius, the total particle
number and the polarization are given by
\begin{eqnarray}
\frac{N}{a^2_xc^2}&=&\int_{-\infty}^{\infty}\tilde{n}(\tilde{x})d\tilde{x}, \nonumber \\
P&=&\int_{-\infty}^{\infty}\tilde{n}_1(\tilde{x})d\tilde{x}/(N/(a^2_xc^2)).
\label{partpol} 
\end{eqnarray}

We extract the  phase boundaries from the equation of state (\ref{pressure}) within the LDA  (\ref{partpol})   in the limit $t\rightarrow 0$, see  Fig. \ref{diagram}.
 The line $R_1$ ($R_2$)  indicates the vanishing density of unpaired bosons (spin-singlet pairs).
The intersection of the   phase boundaries occurs  at $h=0.5$ and $\mu=0.5$ yielding the critical polarization $P_c\approx16.97\%$. 
This phase diagram resembles that of the spin-1/2 attractive Fermi gas, recently confirmed by Liao {\em et al.} \cite{Liao}. 
Here the critical properties of the spin-1 Bose gas are described by two Tonks-Girardeau gases with masses $m$ and $2m$, which are likely to mimic 
the free fermions and  bosonic bound pairs  in the spin-1/2 attractive Fermi gas.

\begin{figure}[t]
{{\includegraphics [width=0.99\linewidth]{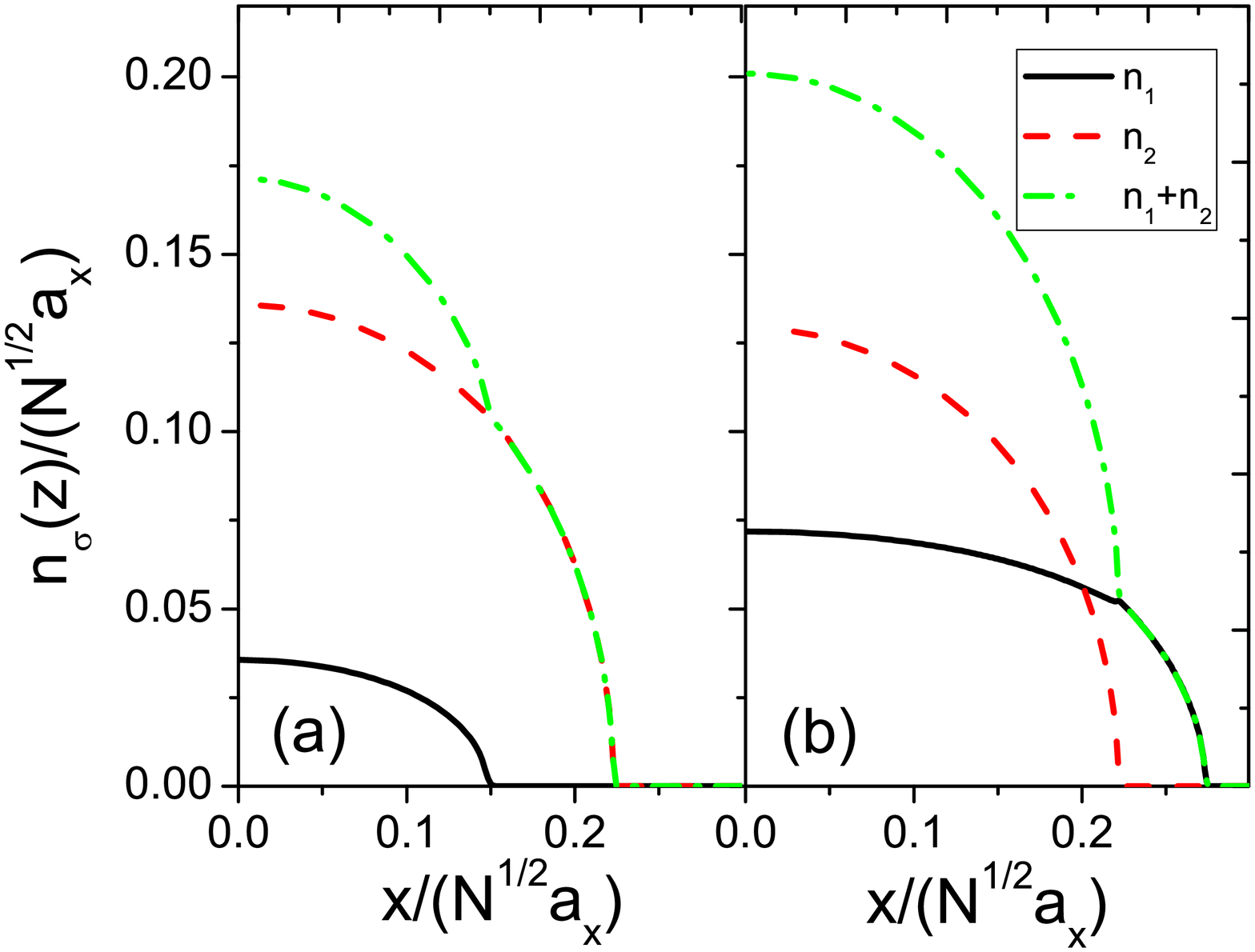}}}
\caption{(Color online) The density distribution profiles for trapped bosons at zero temperature 
with $N/a_x^2c^2\simeq0.05$ for (a) $P\approx12.22\%$ (or $h=0.49$) and (b) $P\approx22.33\%$ 
(or $h=0.51$).} 
\label{denstrap}
\end{figure}

 Fig. \ref{denstrap} shows  the density distribution profiles in 
the limit $t \to  0$ for $N/a_x^2c^2\simeq0.05$  for 
(a) low polarization ($P < P_c$)  $P\approx12.22\%$ and 
(b) large polarization  ($P > P_c$) $P\approx22.33\%$. 
We find a partially polarized core surrounded by two wings composed of either 
  spin-singlet pairs  or a fully spin-aligned  Tonks-Girardeau  gas. The threshold values of the phase boundaries are consistent with the phase diagram 
  Fig.\ref{diagram}.  Our result convincingly indicates that spinor Bose gases
in a 1D harmonic trap constitute an excellent quantum simulator of strongly correlated TLLs  which  could be probed 
by \textit{in situ} imaging, analogously to the trapped Fermi gas \cite{Liao}.

\begin{figure}[t]
{{\includegraphics [width=0.5\linewidth]{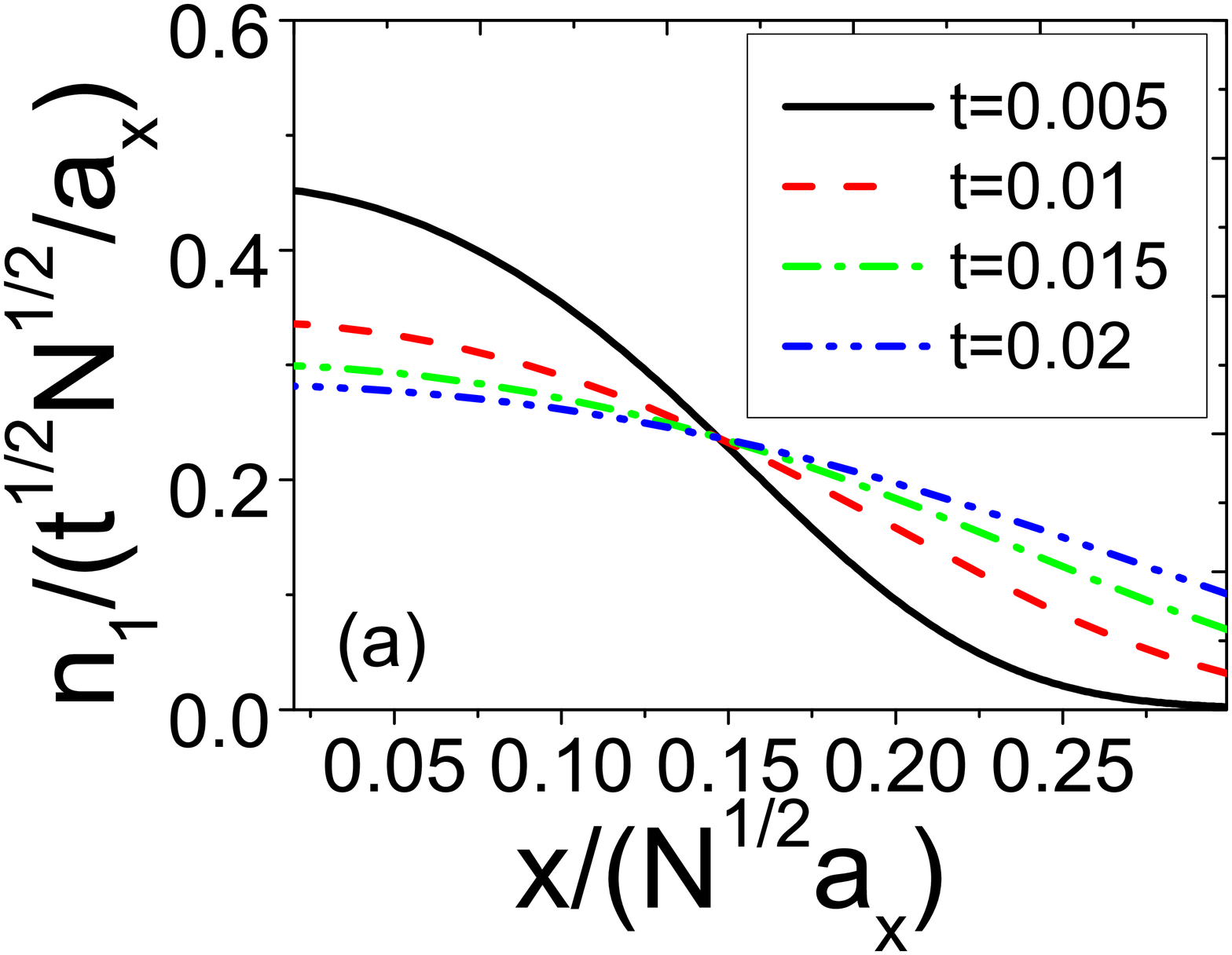}}{\includegraphics [width=0.5\linewidth]{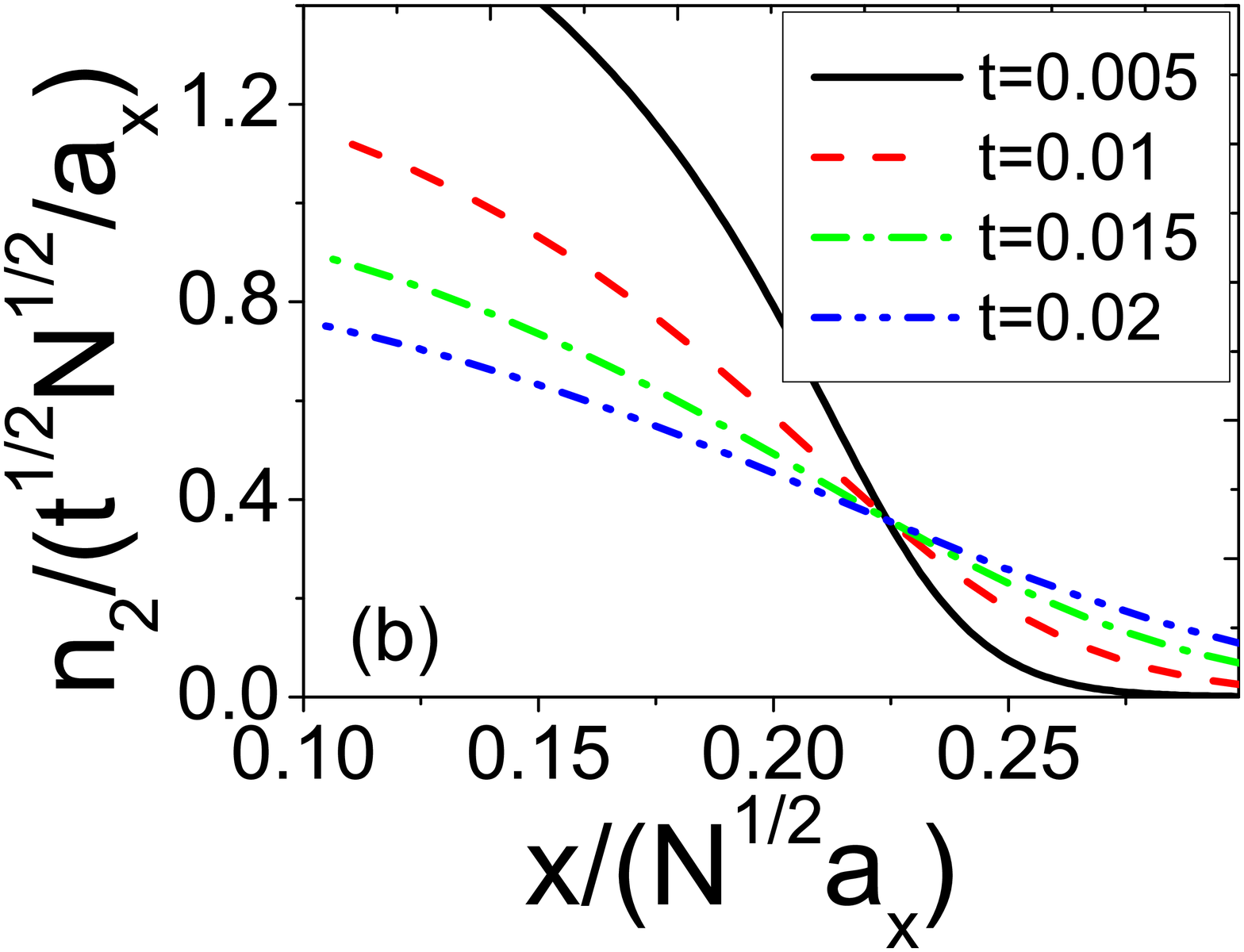}}}\\ 
{{\includegraphics [width=0.5\linewidth]{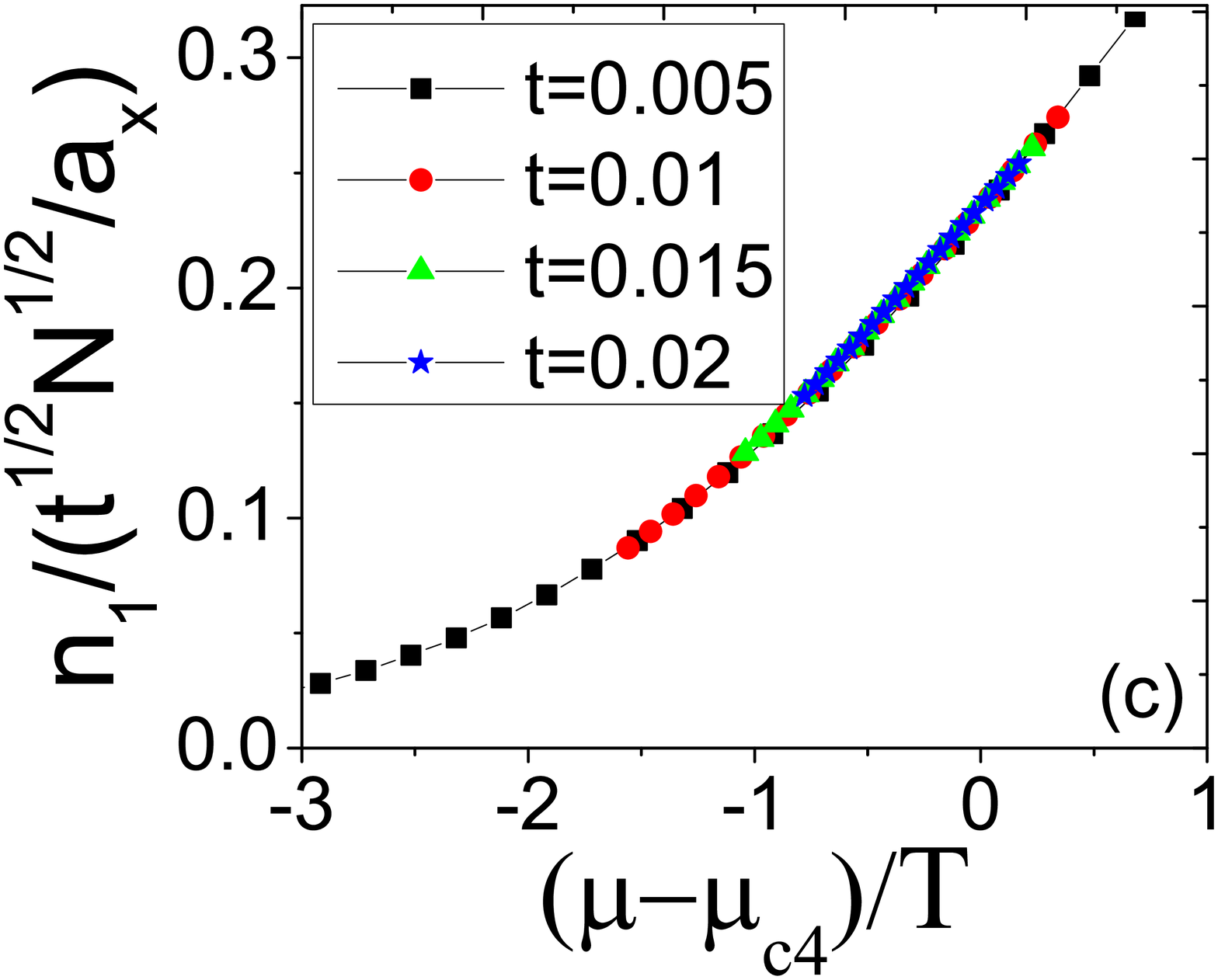}}{\includegraphics [width=0.5\linewidth]{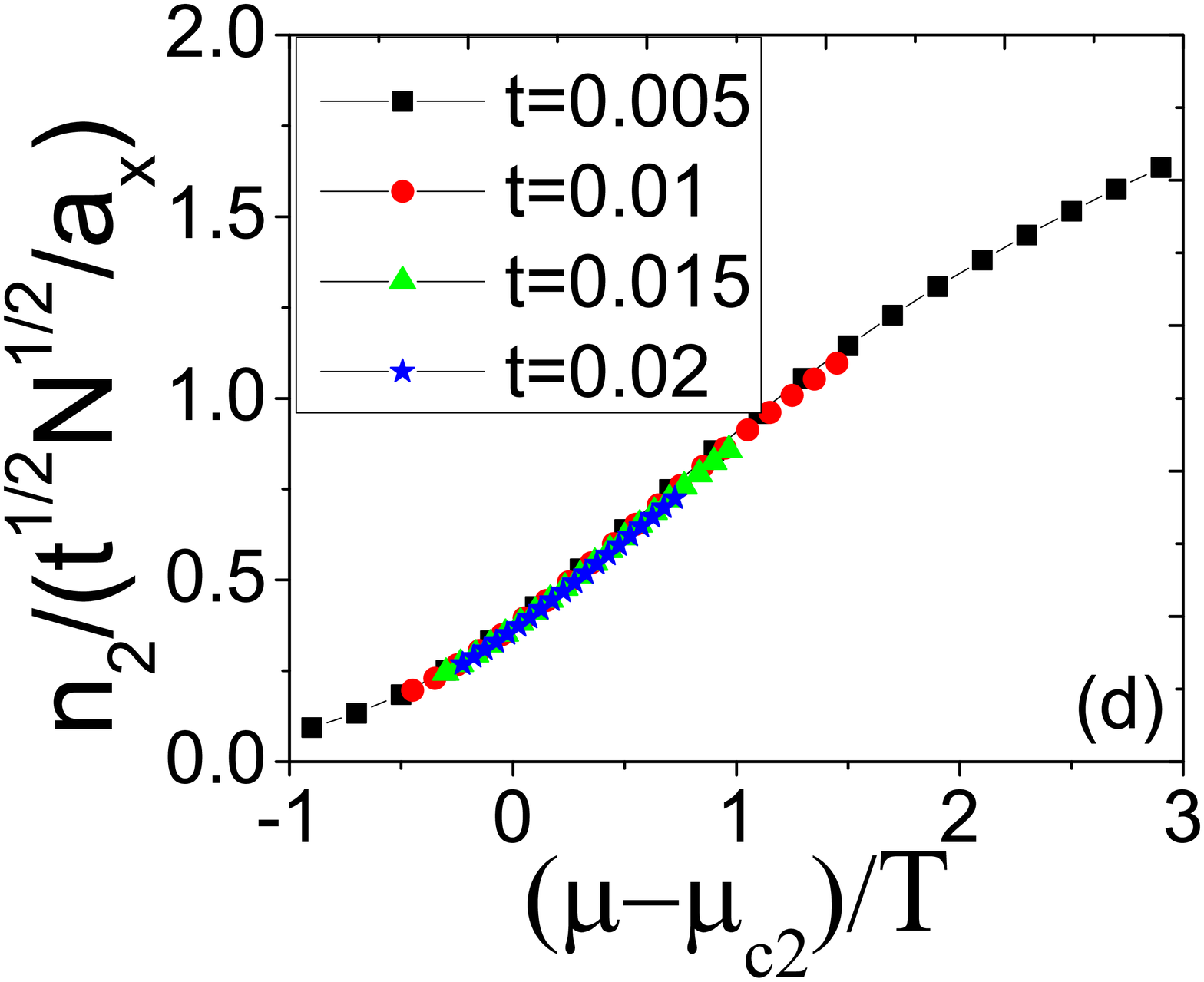}}}
\caption{(Color online) Quantum criticality for low polarization.  
The rescaled density profiles are plotted within LDA near  the critical points (a) $\tilde{\mu}_{c4}$ and  (b) $\tilde{\mu}_{c2}$ 
at different temperatures $t=0.005$, $0.01$, $0.015$ and $0.02$.  All curves intersect at the critical points.
The rescaled density vs (c) $(\mu(x)-\mu_{c4})/T^{z\nu}$ and (d) $(\mu(x)-\mu_{c2})/T^{z\nu}$ at different temperatures $t=0.005$, $0.01$, $0.015$ and $0.02$. 
All data points collapse onto a single curve.}
\label{dens049}
\end{figure}

{\it Quantum criticality} - The equation of state  (\ref{pressure}),  capturing the entire thermal fluctuation in the quantum critical regimes,  
illustrates the microscopic origin of the quantum criticality. The thermodynamic functions can be cast in a 
universal scaling form \cite{Fisher,Sachdev}
For example, the density and compressibility  are given by 
\begin{eqnarray}
n(\mu,T,x)&=&n_0+T^{\frac{d}{z}+1-\frac{1}{\nu z}}{\cal G}\left(\frac{\mu(x)-\mu_c}{T^{\frac{1}{\nu z}}}\right),
\label{uni_scal_dens}\\
\kappa(\mu,T,x)&=&\kappa_0+T^{\frac{d}{z}+1-\frac{2}{\nu z}}{\cal F}\left(\frac{\mu(x)-\mu_c}{T^{\frac{1}{\nu z}}}\right).
\label{uni_scal_kappa}
\end{eqnarray}
with dimensionality $d=1$. 
Here we find  the scaling functions  ${\cal G}(x)=\lambda_{\alpha} {\mathrm{Li}}_{1/2}(-e^{x})$ and ${\cal F}(x)=\lambda_{\beta} {\mathrm{Li}}_{-1/2}(-e^{x})$, 
from which follow the dynamical critical 
exponent $z=2$ and correlation length exponent $\nu =1/2$ for different phases of the spin states.  
In the above,  $n_0$, $\kappa_0$,   $\lambda_{\alpha}$ and $\lambda_{\beta}$ are  constants associated with the background near the 
critical points (see details in \cite{kuhn}). Moreover,  quantum criticality driven by the external field $H$ leads to similar scaling, 
where now $ {\cal G}(x)=\lambda_{g} {\mathrm{Li}}_{1/2}(-e^{\Lambda_0 x})$ and ${\cal F}(x)=\lambda_{f} {\mathrm{Li}}_{-1/2}(-e^{\Lambda_0 x})$ with a dimensionless 
factor $\Lambda_0$ and  the same critical exponents \cite{kuhn}. Here $\lambda_{g,f}$ are  different  constants associated with  critical points.

The critical exponents appearing in the scaling functions are analytically determined from the exact Bethe ansatz solution.
We observe that the 1D spin-1 Bose gas belongs to the same universality class  as spin-1/2 attractive fermions \cite{Guan} due to the hard-core 
nature of the two coupled Tonks-Girardeau gases. 
This demonstrates a significant  feature of 1D many-body physics -- quantum criticality involves a universal crossover from  
a TLL with linear dispersion to free fermions with a quadratic dispersion in the  low energy physics \cite{Masaki}.  
In addition, the critical exponents of 1D spin-1 bosons have the
same values as the critical exponents of the Mott transition in 1D and 2D 
Bose-Hubbard models, see \cite{Campostrini2,Zhang:2012}.

The universal scaling behaviour of the homogenous system can be mapped out through the density profiles of the trapped gas 
at finite temperatures. Although a finite-size scaling effect is evident in the scaling analysis  \cite{Campostrini,Zhou-Ho,Ceccarelli}, 
the finite-size error lies within the current experimental  accuracy \cite{Zhang:2012}. 
Here we show  that  all thermodynamic observables are given in terms of  the rescaled  temperature and chemical potential. 
Thus one can either lower the temperature or increase the interaction strength (by tuning the transverse frequency) such that all 
data curves for the physical properties  in the trapped gas collapse at different temperatures into a single curve with a proper scaling. 
This collapse signature can  be used to confirm the quantum critical law, see the experimental measurements given in \cite{Zhang:2012}. 

 In the regime of low polarization ($P < P_c$),  the phase transitions   from the
vacuum into the spin-singlet paired  phase and from the pure paired phase into the 
mixture of spin-singlet pairs and  spin-aligned bosons  occur as   the chemical potential passes the lower critical point 
 $\tilde{\mu}_{c2}=-0.5$   and the upper critical point 
$\tilde{\mu}_{c4}= -h + \frac{32\sqrt{2}}{15\pi}\left(\frac{1}{2}-h\right)^{\frac{3}{2}}
 + \frac{2912}{225\pi^2}\left(\frac{1}{2}-h\right)^2$, respectively.

 The universal scaling behaviour can be identified  following the scheme  proposed in \cite{Zhou-Ho}  by  plotting 
 the ``rescaled density'' versus the chemical potential for different values of temperature. 
Figures \ref{dens049} (a) and (b) show the
rescaled density normalized by $N^{1/2}/(a_x)$ versus the rescaled distance  at different temperatures. 
All curves intersect  at the  critical points  $\tilde{\mu}_{c2}$ and $\tilde{\mu}_{c4}$ after a  subtraction of the background density $n_0$. 
We also see in Fig. \ref{dens049} that all data points collapse onto a single curve near  (c) $\tilde{\mu}_{c4}$ and (d) $\tilde{\mu}_{c2}$, 
which is the scaling function for these quantum critical points.

\begin{figure}[t]
{{\includegraphics [width=0.5\linewidth]{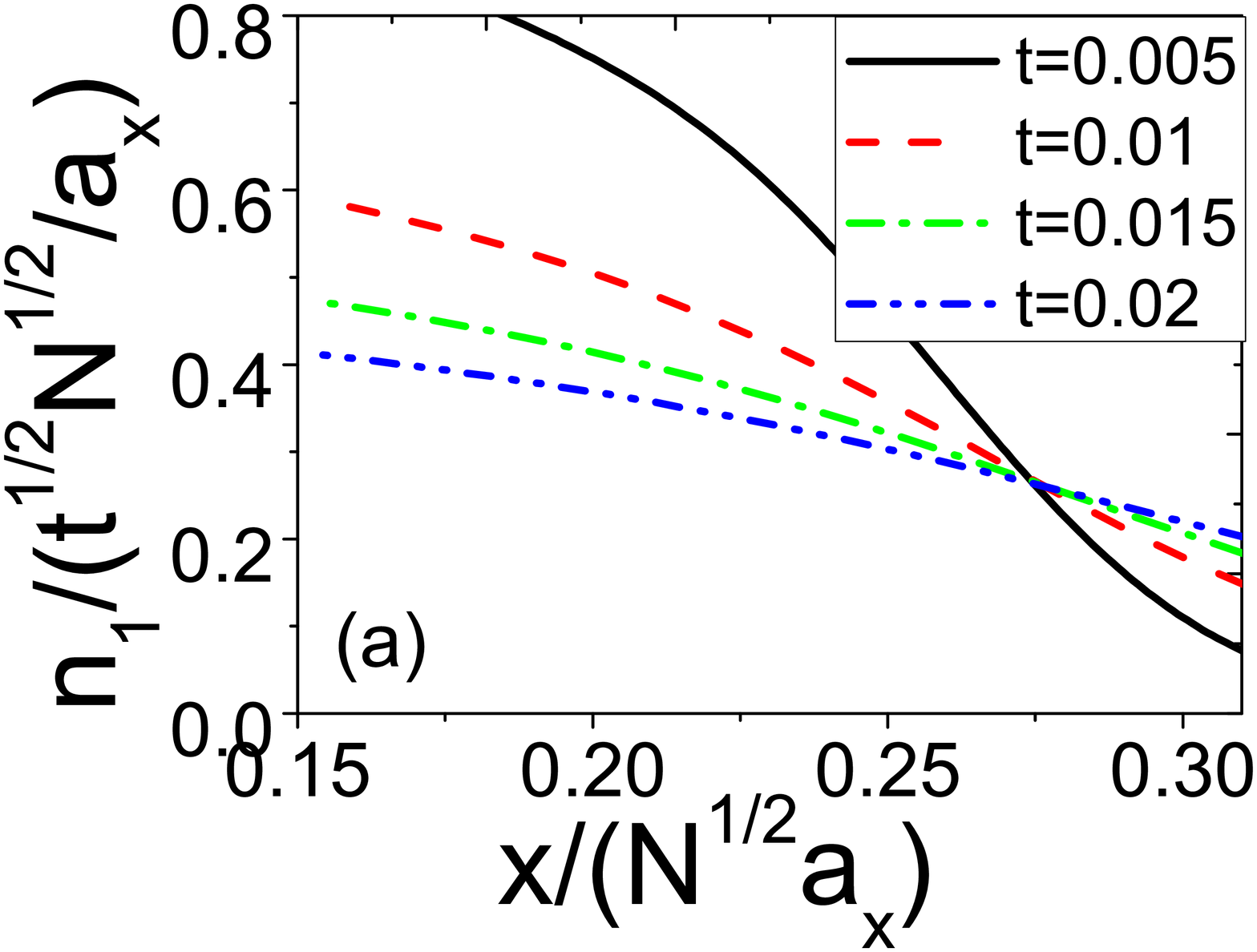}}{\includegraphics [width=0.5\linewidth]{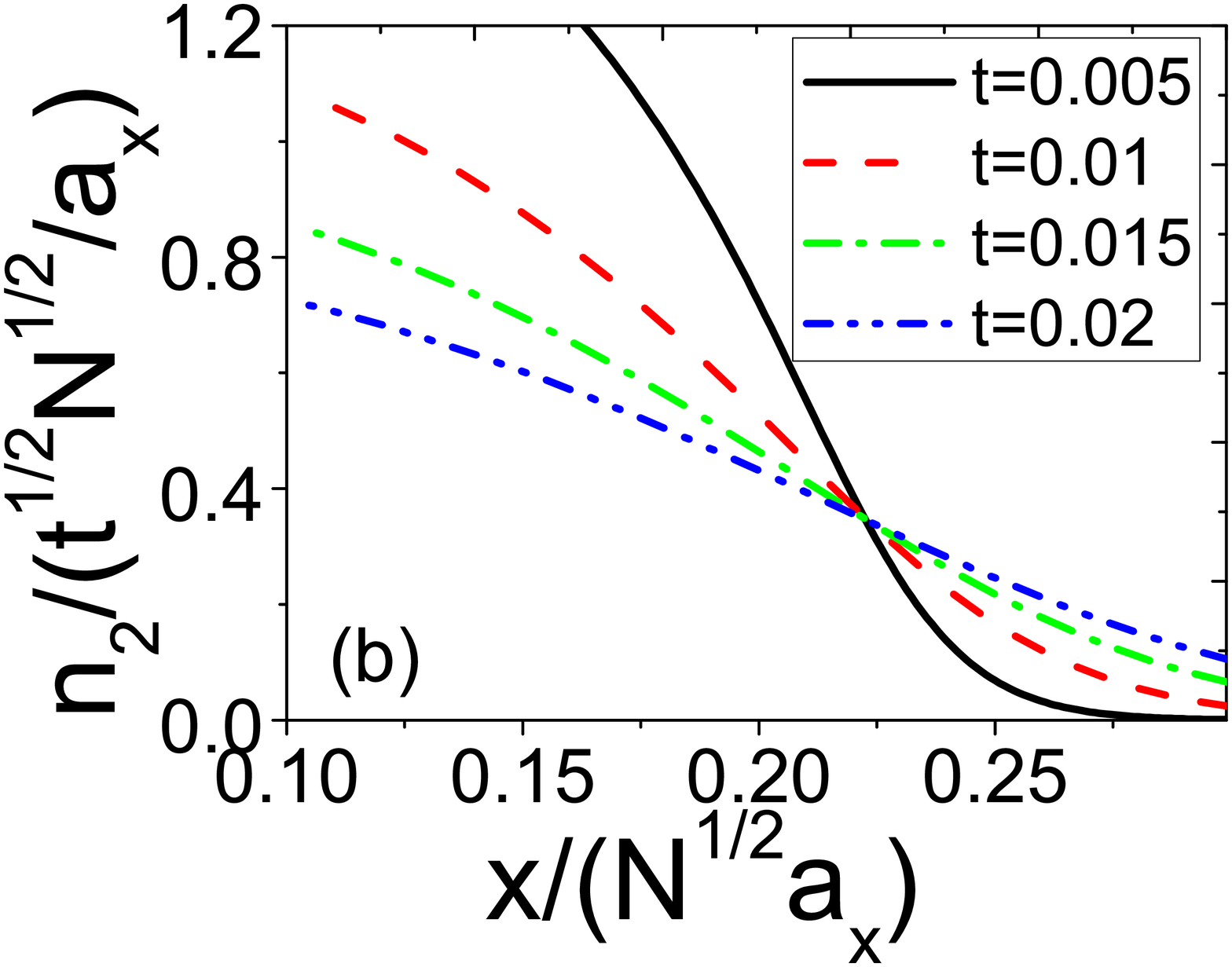}}}\\ 
{{\includegraphics [width=0.5\linewidth]{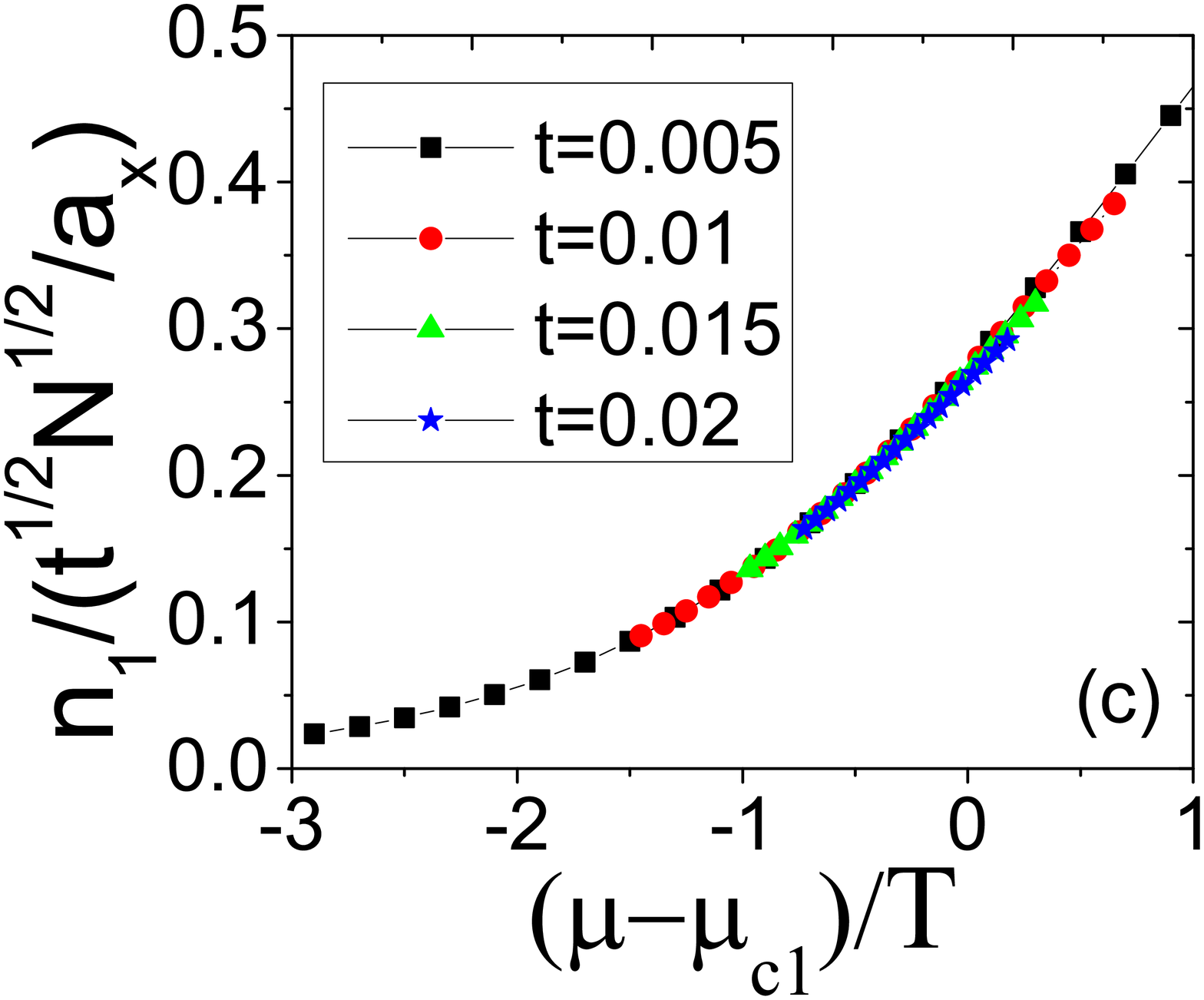}}{\includegraphics [width=0.5\linewidth]{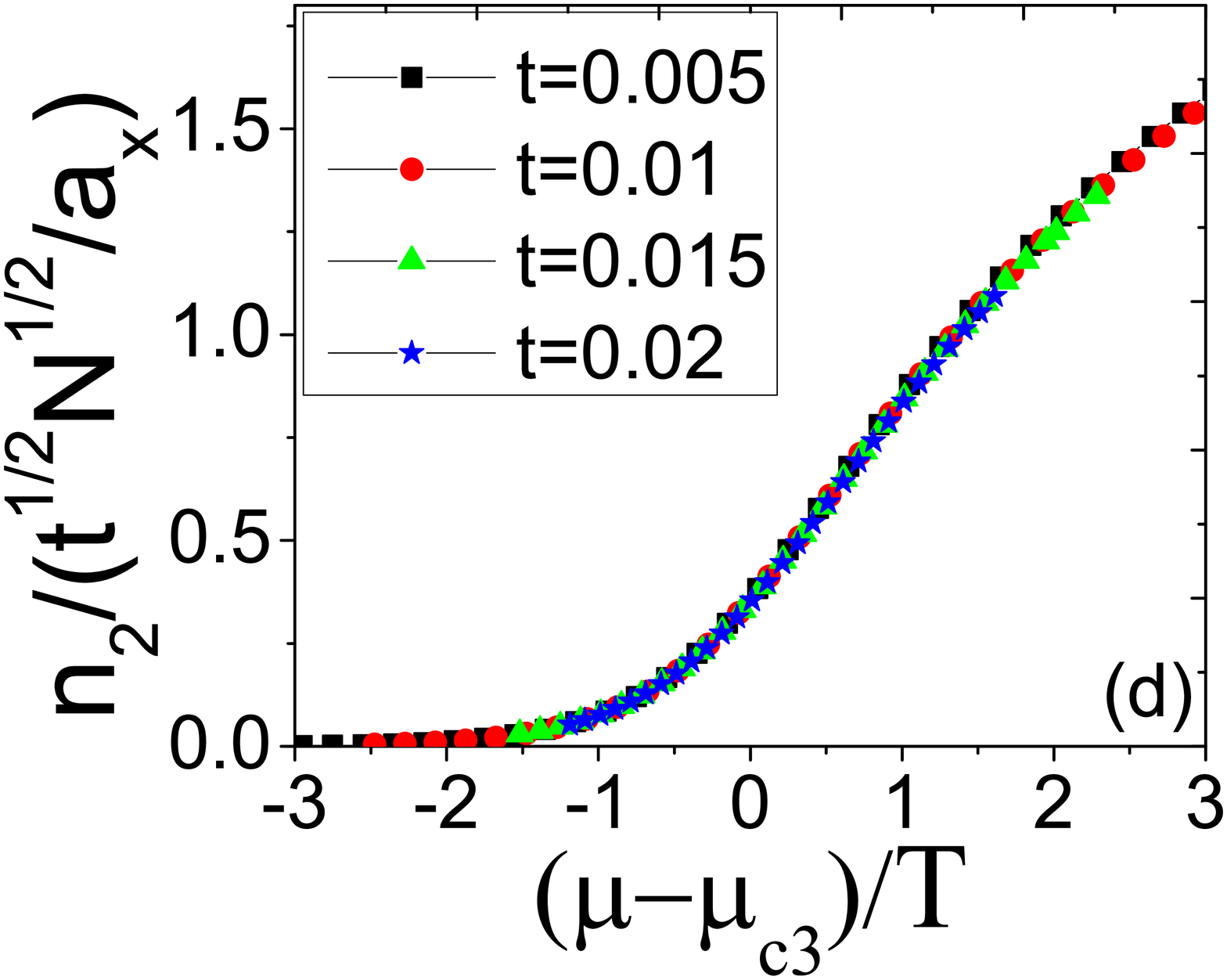}}}
\caption{(Color online) Quantum criticality for high polarization. 
The rescaled density profiles are plotted within LDA near the critical points (a) $\tilde{\mu}_{c1}$ and  (b) $\tilde{\mu}_{c3}$. 
The rescaled density vs (c) $(\mu(x)-\mu_{c1})/T^{z\nu}$ and (d) $(\mu(x)-\mu_{c3})/T^{z\nu}$. As in Fig. \ref{dens049} the 
curves and data points are indicative of scaling behaviour at quantum criticality.}
\label{dens051}
\end{figure}

For large polarization ($P>P_c$), the phase transitions  from vacuum into the  ferromagnetic spin-aligned  
boson phase and from the spin-aligned boson phase into 
the mixture of spin-singlet pairs and  spin-aligned bosons  occur as   the chemical potential varies across  the lower critical point 
$\tilde{\mu}_{c1}=-h$  and the  upper critical point 
$\tilde{\mu}_{c3}= -\frac{1}{2} + \frac{8\sqrt{2}}{15\pi}\left(h-\frac{1}{2}\right)^{\frac{3}{2}} 
+ \frac{104}{75\pi^2}\left(h-\frac{1}{2}\right)^2$, respectively.  A similar  analysis is presented 
in Fig.~\ref{dens051} near the critical points $\tilde{\mu}_{c1}$ and $\tilde{\mu}_{c3}$.

Similar plots can be made for the rescaled compressibility in agreement with the universal scaling result (\ref{uni_scal_kappa}).
The compressibility curves at different temperatures collapse into a single curve obeying the scaling law. 
The compressibility always develops a peak near  the phase boundary on the side with higher density of state. 
However, the experimental measurement of the compressibility is more difficult than the density profiles. 
Nevertheless, progress in measuring the compressibility of an ultracold Fermi gas has recently been made \cite{compress}.


In conclusion, we have discussed the equation of state, the density profiles, TLL thermodynamics and universal scaling 
at quantum criticality for the spin-1 Bose gas with strongly repulsive density-density and antiferromagnetic spin-exchange 
interactions in a 1D harmonic trap. We have shown that the quantum criticality of different spin states belongs to the universality 
class with critical exponents $z=2$ and $\nu=1/2$. This remains true in general for the 1D model even for the non-integrable case 
$c_0 \neq c_2$. We have also demonstrated that the phase diagram, the TLLs and critical properties of the bulk system can be 
mapped out from the density profiles of the trapped spinor gas at finite temperatures. Current experiments \cite{Exp10, Exp11, Liao, armijo} 
are capable of measuring such universal features of 1D many-body physics.

{\bf Acknowledgements}- This work has been partially supported by the Australian Research Council. 
CCNK thanks CAPES (Coordenacao de Aperfeicoamento de pessoal de Nivel Superior) for financial support.
He also thanks the Department of Theoretical Physics, RSPE at  ANU for their hospitality. 
AF thanks CNPq (Conselho Nacional de Desenvolvimento Cientifico e Tecnol\'ogico) for financial support.

\end{document}